\documentclass{andp2012}
\usepackage[english]{babel}
\usepackage{epsfig}
\usepackage{graphicx}
\title{Quantum thermalization and thermal entanglement in the open quantum Rabi model}
\author[W.-Y. Liu]{Wang-Yan Liu\inst{1}}
\author[L.-B. Fan]{Li-Bao Fan\inst{1,2}}
\author[Y.-X. Zeng]{Ye-Xiong Zeng\inst{3}}
\author[J.-F. Huang]{Jin-Feng Huang\inst{1,}\footnote{Corresponding author\quad E-mail:~\textsf{jfhuang@hunnu.edu.cn}}}
\author[J.-Q. Liao]{ Jie-Qiao Liao\inst{1,}\footnote{Corresponding author\quad E-mail:~\textsf{jqliao@hunnu.edu.cn}}}
\address[1]{Key Laboratory of Low-Dimensional Quantum Structures and Quantum Control of
Ministry of Education, Key Laboratory for Matter Microstructure and Function of Hunan Province, Department of Physics and Synergetic Innovation Center for Quantum Effects and Applications, Hunan Normal University, Changsha 410081, China}
\address[2]{Hunan Key Laboratory of Nanophotonics and Devices,Hunan Key Laboratory of Super-Microstructure and Ultrafast Process,School of Physics and Electronics, Central South University, Changsha 410083, China}
\address[3]{Theoretical Quantum Physics Laboratory, RIKEN Cluster for Pioneering Research, Wako-shi, Saitama 351-0198, Japan}
\shortauthors{F. Author et al.}
\begin{abstract}
We study quantum thermalization and thermal entanglement in the open
quantum Rabi model (QRM), in which a two-level system and a single-mode bosonic field are coupled to either two individual heat baths or a common heat bath. By treating the QRM as an effective multilevel system and deriving global quantum master equations in the eigenstate representation of the QRM, we study the physical conditions
for quantum thermalization of the QRM. It is found that, in the individual heat-bath case, the QRM can only be thermalized when either the two heat baths have the same temperature or the QRM is only coupled to one of the two baths. In the common heat-bath case, differently, the QRM can always be thermalized. We also study thermal entanglement of the QRM in both the resonant- and non-resonant coupling cases. The logarithmic negativity for the thermal state of the QRM is obtained in a wide parameter space, ranging from the low- to high-temperature limits, and from the weak- to deep-strong-coupling regimes. This work paves the way towards the study of quantum effects in nonequilibrium ultrastrongly-coupled light-matter systems.
\end{abstract}
\shortabstract
\begin{document}
\captionsetup[figure]{labelfont={bf},labelformat={default},labelsep=period,name={Figure}}
\captionsetup[Table]{labelfont={bf},labelformat={default},labelsep=period,name={Table}}
\maketitle

\section{Introduction}

Quantum thermalization,\textsuperscript{\cite{bib1,bib2}} as one of the most important topics in quantum thermodynamics, is understood as an irreversible dynamic process via which a quantum
system immersed in its environment approaches a thermal equilibrium state at
the same temperature as the environment. Until now, considerable studies have been done on both quantum thermalization of coupled quantum systems and thermal entanglement\textsuperscript{\cite{bib5}} between the subsystems in coupled systems. For instance, quantum thermalization and thermal entanglement of two coupled two-level atoms have been studied in the dressed-state representation.\textsuperscript{\cite{bib6}} Recently, quantum thermalization and thermal entanglement
in the open Jaynes-Cummings model (JCM) have also been studied.\textsuperscript{\cite{bib7}} In addition, thermal entanglement has been studied in various physical systems, including coupled spins\textsuperscript{\cite{bib8,bib9,bib10,bib11,bib12}} and coupled oscillators.\textsuperscript{\cite{bib13}}

In general, the thermal state of a coupled quantum system should be entangled because the thermal state takes the form as $\rho_{\mathrm{th}}=Z_{\mathrm{sys}}^{-1}\exp({-\beta H_{\mathrm{sys}}})$ with $H_{\mathrm{sys}}=H_{A}+H_{B}+H_{I}$, where $H_{A}$ and $H_{B}$ are, respectively, the free Hamiltonians of the subsystems $A$ and $B$, $H_{I}$ is the interaction Hamiltonian between $A$ and $B$, and $Z_{\mathrm{sys}}=\mathrm{Tr}[\exp({-\beta H_{\mathrm{sys}}})]$ is the partition function with $\beta=1/(k_{B}T)$ being the inverse temperature ($k_{B}$  is the Boltzmann constant). However, a counterintuitive phenomenon of vanishing thermal entanglement in the open JCM\textsuperscript{\cite{bib7}} has been
found and proved. Usually, the JCM\textsuperscript{\cite{bib14}} is obtained by discarding the counter-rotating terms in the quantum Rabi model (QRM)\textsuperscript{\cite{bib15}} with the rotating-wave approximation (RWA), when the interaction between a two-level system (TLS) and a single-mode bosonic field does not enter the ultrastrong-coupling regime\textsuperscript{\cite{bib16}} Therefore, a natural and important question is what happens with the thermalization and thermal entanglement of the open QRM in the ultrastrong-coupling regime. In particular, great advances have been made in the enhancement of the coupling strength of the QRM in the last decade, and the ultrastrong couplings\textsuperscript{\cite{bib17,bib18,bib19}} even deep-strong couplings\textsuperscript{\cite{bib20}} in the QRM have been realized in various physical systems, such as superconducting quantum circuits\textsuperscript{\cite{bib19,bib21,bib22,bib23,bib24}}
and semiconductor quantum wells.\textsuperscript{\cite{bib25,bib26,bib27}} Therefore, the study of quantum thermalization and thermal entanglement in the QRM becomes an urgent topic to be addressed.

In this paper, we study quantum thermalization of the open QRM, which is coupled to either two individual heat baths (IHBs) or
one common heat bath (CHB). Concretely, we derive global quantum master equations\textsuperscript{\cite{bib28,bib29,bib30,bib31,bib32,bib33}} to describe the evolution of the QRM. The global quantum master equations are valid even when the interaction between the TLS and the bosonic mode is much stronger than the system-bath couplings. Note that quantum statistics based on the global quantum master equations have been studied in coupled atom systems\textsuperscript{\cite{bib34,bib35,bib36,bib37,bib38,bib39,bib40,bib41,bib42,bib43,bib44,bib45,bib46,bib47}} and coupled harmonic-oscillator systems.\textsuperscript{\cite{bib48,bib49,bib50}} To characterize the thermalization of the QRM, we introduce the effective temperatures associated with any two eigenstates, and evaluate the thermalization of the QRM by inspecting whether its steady-state density matrix can be expressed
as a thermal equilibrium state. Furthermore, we study
thermal entanglement\textsuperscript{\cite{bib8}} between the two subsystems of the QRM by calculating
the logarithmic negativity of the thermal state.
We obtain the thermal entanglement of the QRM in both the resonant- and nonresonant-coupling cases. Concretely, we find the dependence of the logarithmic negativity on the coupling strength and the temperature. We also explain the thermal entanglement in the low-temperature limit based on the eigenstate entanglement.

The rest of this paper is organized as follows. In Sec.~\ref{II}, we introduce
the open QRM in both the resonant- and nonresonant-coupling cases, and present the Hamiltonians of the system. In Secs.~\ref{III} and~\ref{IV},
we study quantum thermalization of the QRM in both the IHB and CHB cases,
respectively. In Sec.~\ref{V}, the thermal entanglement of the QRM is analyzed in both the resonant- and nonresonant-coupling cases.
A conclusion is given in Sec.~\ref{VI}.

\begin{figure}
	\begin{centering}
		\includegraphics[width=0.46\textwidth]{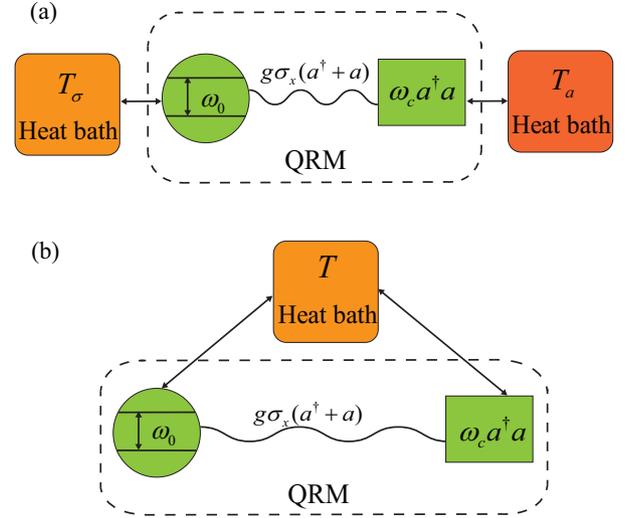}
		\caption{Schematic of the open QRM in either (a) the individual heat-bath case
			or (b) the common heat-bath case.}
		\label{Fig1}
		\par\end{centering}
\end{figure}

\section{Model and Hamiltonian}\label{II}

We consider a quantum Rabi model which describes the interaction of a TLS with a single-mode bosonic field through the dipole coupling (see Fig.~\ref{Fig1}).
The two levels of the TLS are denoted as the ground state $\vert g\rangle$
and the excited state $\vert e\rangle$, and the energy separation between
these two levels is $\hbar$$\omega_{0}$. We introduce the Pauli operators
$\sigma_{x}=\vert e\rangle\langle g\vert+\vert g\rangle\langle e\vert$,
$\sigma_{y}=i(\vert g\rangle\langle e\vert-\vert e\rangle\langle g\vert)$,
and $\sigma_{z}=\vert e\rangle\langle e\vert-\vert g\rangle\langle g\vert$
to describe the TLS. For the single-mode bosonic field, we assume its resonance frequency as $\omega_{c}$ and denote its annihilation (creation) operator as $a$  $(a^{\dagger})$. The Hamiltonian of the QRM takes the form as

\begin{equation}
	H_{\mathrm{QRM}}=\frac{\hbar\omega_{0}}{2}\sigma_{z}+\hbar\omega_{c}a^{\dagger}a+\hbar g\sigma_{x}(a^{\dagger}+a).
\end{equation}

Here, the first and second terms describe the free Hamiltonian of the TLS and the single-mode bosonic field, respectively. The last term describes the interaction between the TLS and the field, with $g$ being the coupling strength. When $g/\omega_{0},~\omega_{c}>0.1$, the QRM enters the ultrastrong-coupling regime. The QRM is in the deep-strong-coupling regime when $g/\omega_{0},~\omega_{c}>1$.

For the QRM, the total excitation operator $N=a^{\dagger}a+\vert e\rangle\langle e\vert$ is no longer a conserved quantity. However, there exists a parity operator
$P=-\sigma_{z}(-1)$$^{a^{\dagger}a}$, which commutes with the Rabi Hamiltonian $H_{\mathrm{QRM}}$, and hence we can divide the whole Hilbert space of the system into odd and even parity subspaces.\textsuperscript{\cite{bib51,bib52}} Generally, it is hard to analytically solve the eigensystem of the QRM with the element functions. However, the eigenenergy spectrum of the QRM has been obtained
with several methods under the assistant of numerical calculations.\textsuperscript{\cite{bib15,bib53,bib54,bib55}}
In this work, we will numerically solve the eigensystem of the QRM and study the steady-state populations of these eigenstates in a sufficiently large Hilbert space.
\begin{figure}[h]
	\begin{centering}
		\includegraphics[width=0.48\textwidth]{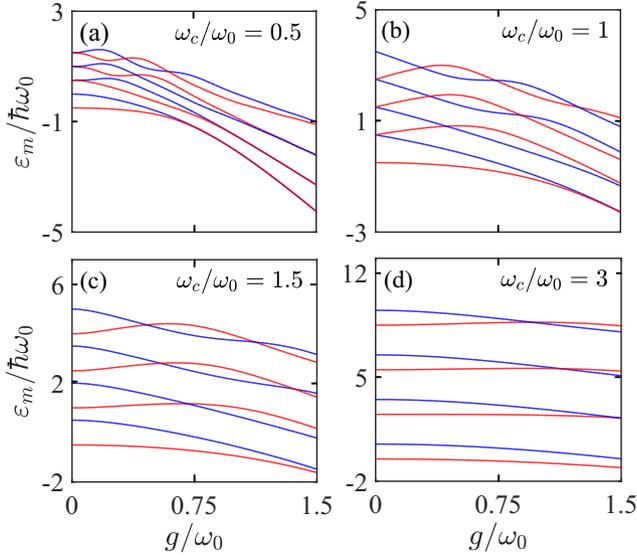}
		\caption{Scaled energies $\varepsilon_{m}/\hbar\omega_{0}$ for the lowest eight eigenstates as functions of $g/\omega_{0}$ at different values of $\omega_{c}/\omega_{0}$: (a) $\omega_{c}/\omega_{0}=0.5$, (b) $\omega_{c}/\omega_{0}=1$, (c) $\omega_{c}/\omega_{0}=1.5$, and (d) $\omega_{c}/\omega_{0}=3$. Here, the eigenstates with even and odd parities are plotted with red and blue curves, respectively.}
		\label{Fig14}
		\par\end{centering}	
\end{figure}
For expressional convenience, we denote the eigensystem of the Rabi Hamiltonian as
$H_{\mathrm{QRM}}\vert\varepsilon_{m}\rangle=\varepsilon_{m}\vert\varepsilon_{m}\rangle$ ($m=1$, $2$, $3$, ..., $\infty$),
where $\vert\varepsilon_{m}\rangle$ are eigenstates with the corresponding
eigenvalues $\varepsilon_{m}$. These eigenstates can be expanded with the bare vectors $\vert e,m\rangle$ and $\vert g,n\rangle$
	as $\vert\varepsilon_{m}\rangle =\sum_{n=0}^{\infty}C_{g,2n}\vert g,2n\rangle+\sum_{m=0}^{\infty}C_{e,2m+1}\vert e,2m+1\rangle$ and $\vert\varepsilon_{m}\rangle =\sum_{n=0}^{\infty}C_{g,2n+1}\vert g,2n+1\rangle+\sum_{m=0}^{\infty}C_{e,2n}\vert e,2n\rangle$ for even and odd parities respectively, where the superposition coefficients can be determined numerically.

In Fig.~\ref{Fig14}, we show the lowest eight eigenstate energies of the QRM as a function of the coupling strength in both resonant and nonresonant-coupling cases. We can see several features in the energy spectrum. (i) The entire pattern of these levels decreases with the increase of the coupling strength $g/\omega_{0}$. For a larger value of the ratio $\omega_{c}/\omega_{0}$, the slope is smaller.
(ii) These energy levels can be paired off from lower states to upper states. Except for the first pair of states, the two states in other pairs intersect with each other, and the crossing levels have different parities. With the increase of the ratio $\omega_{c}/\omega_{0}$, the horizontal distance between two neighboring crossing points increases. (iii) In the absence of the coupling, i.e., $g=0$, these states are reduced to the bare states, and hence the states with the same excitation are degenerate in the resonant case [i.e., panel (b)]. (iv) When the system enters the deep-strong coupling regime, the states in the same pair become near degenerate. Note that the information of the energy spectrum will be useful
for analyzing the thermal entanglement.

To characterize the interactions of both the TLS and the bosonic mode with the environments,
we consider two different cases: (i) The TLS
and the bosonic mode are coupled to individual heat bath alone [Fig.~\ref{Fig1}(a)]. (ii) Both the TLS and the bosonic mode are coupled to a common heat bath [Fig.~\ref{Fig1}(b)]. In the IHB case, the free Hamiltonians of the two heat baths are expressed as $H_{B}^{(\mathrm{IHB})}=\sum_{q}\hbar\omega_{q}C_{q}^{\dagger}C_{q}+\sum_{k}\hbar\omega_{k}A_{k}^{\dagger}A_{k}$,
where the creation and annihilation operators $C_{q}^{\dagger}$ $ (A_{k}^{\dagger})$
and $C_{q}$ $ (A_{k})$ describe the $q$th ($k$th) mode with
frequency $\omega_{q}$ $(\omega_{k})$ in the heat bath of the TLS (bosonic
mode). The interaction Hamiltonian between the QRM and the two heat baths
can be expressed as

\begin{equation}
	H_{I}^{\text{(}\mathrm{IHB})}=\sum_{q}\hbar\lambda_{q}\sigma_{x}(C_{q}^{\dagger}+C_{q})+\sum_{k}\hbar\eta_{k}(a^{\dagger}+a)(A_{k}^{\dagger}+A_{k}),
\label{HI1}
\end{equation}
where $\lambda_{q}$ $(\eta_{k})$ is the coupling strength between the
TLS (bosonic mode) and the $q$th ($k$th) mode of its heat bath.

In the CHB case, both the TLS and the bosonic mode are connected with a
common heat bath described by the Hamiltonian $H_{B}^{\text{(}\mathrm{CHB})}=\sum_{p}\hbar\omega_{p}B_{p}^{\dagger}B_{p}$,
where $B_{p}^{\dagger}$ and $B_{p}$ are, respectively, the creation
and annihilation operators of the $p$th mode with resonance frequency $\omega_{p}$ in the common bath. The interaction Hamiltonian
between the QRM and the CHB reads
\begin{equation}
	H_{I}^{\text{(}\mathrm{CHB})}=\sum_{p}\hbar\lambda_{p}\sigma_{x}(B_{p}^{\dagger}+B_{p})+\sum_{p}\hbar\eta_{p}(a^{\dagger}+a)(B_{p}^{\dagger}+B_{p}),
	\label{3}
\end{equation}
where $\lambda_{p}$ $(\eta_{p})$ is the coupling strength between the
TLS (bosonic mode) and the $p$th mode of the CHB.

In both the IHB and CHB cases, the Hamiltonian of the whole system including the QRM and the heat baths
can be written as $H=H_{\mathrm{QRM}}+H_{B}^{(s)}+H_{I}^{(s)}$.
Here, $H_{\mathrm{QRM}}$ is the Rabi Hamiltonian, $H_{B}^{(s)}$ and
$H_{I}^{(s)}$ are, respectively, the bath Hamiltonians and the interaction Hamiltonians
between the QRM and its baths, with $s=\mathrm{IHB}$ and $\mathrm{CHB}$ corresponding to the individual heat-bath and common heat-bath cases, respectively.

\section{Quantum Thermalization of the QRM in the IHB Case}\label{III}

In this section, we study the quantum thermalization of the open QRM
in the IHB case. We derive a global quantum master equation
to govern the evolution of the QRM. We also investigate the effective temperature associated with any two eigenstates
to evaluate the quantum thermalization of the QRM.

\subsection{Global quantum master equation}

To include the dissipation in this system, we derive the quantum master equation within the Born--Markov framework, which is valid under the assumption of weak
system-bath coupling and short bath correlation time. In particular, the quantum master equation is derived in the eigenstate representation
of the Rabi Hamiltonian. In the interaction picture with respect to $H_{0}=H_{\mathrm{QRM}}+H_{B}^{(\mathrm{IHB})}$, the master equation can be written as\textsuperscript{\cite{bib2a,bib3,bib4}}
\begin{equation}
		\frac{d}{dt}\tilde{\rho}_{S}(t)=-\int_{0}^{\infty}ds\mathrm{Tr}_{B}[V_{I}^{(\mathrm{IHB})}(t),[V_{I}^{(\mathrm{IHB})}(t-s),\tilde{\rho}_{S}(t)\otimes\rho_{B}]],
		\label{4}
\end{equation}
where $V_{I}^{(\mathrm{IHB})}(t)=\exp(iH_{0}t/\hbar)H_{I}^{(\mathrm{IHB})}\exp(-iH_{0}t/\hbar)$ is the interacting Hamiltonian in the interaction picture. Based on Eqs. (\ref{HI1}), (\ref{4}) and the nonzero correlation functions
\begin{eqnarray}
Tr_{B}\left[A_{k\prime}A_{k}^{\dagger}\rho_{B}\right] & = & \left[\bar{n}_{a}\left(\omega_{k}+1\right)\right]\delta_{k^{\prime},k},\nonumber \\
Tr_{B}\left[A_{k\prime}^{\dagger}A_{k}\rho_{B}\right] & = & \bar{n}_{a}\left(\omega_{k}\right)\delta_{k^{\prime},k},\nonumber \\
Tr_{B}\left[C_{q\prime}C_{q}^{\dagger}\rho_{B}\right] & = & \left[\bar{n}_{\sigma}\left(\omega_{q}+1\right)\right]\delta_{q^{\prime},q},\nonumber \\
Tr_{B}\left[C_{q\prime}^{\dagger}C_{q}\rho_{B}\right] & = & \left[\bar{n}_{\sigma}\left(\omega_{q}\right)\right]\delta_{q^{\prime},q},
\end{eqnarray}
with the average excitation defined below, we can derive the
global quantum master equation of the system in the IHB case as
\begin{align}
	\frac{d}{dt}\tilde{\rho}_{S}(t)= & L_{\mathrm{IHB}}[\tilde{\rho}_{S}(t)]=\sum_{m,n=1,m>n}^{\infty}\sum_{l=\sigma,a}\frac{1}{2}\gamma_{l}(\omega_{m,n})\nonumber \\
	& \times\vert\chi_{l,m,n}\vert^{2}[\overline{n}_{l}(\omega_{m,n})+1]D[\vert\varepsilon_{n}\rangle\langle\varepsilon_{m}\vert]\tilde{\rho}_{S}(t)\nonumber \\
	& +\sum_{m,n=1,m>n}^{\infty}\sum_{l=\sigma,a}\frac{1}{2}\gamma_{l}(\omega_{m,n})\nonumber \\
	& \times\vert\chi_{l,m,n}\vert^{2}\overline{n}_{l}(\omega_{m,n}) D[\vert\varepsilon_{m}\rangle\langle\varepsilon_{n}\vert]\tilde{\rho}_{S}(t),
	\label{5}
\end{align}
where the Lindblad superoperators $D[\vert\varepsilon_{n}\rangle\langle\varepsilon_{m}\vert]\tilde{\rho}_{S}(t)$
and $D[\vert\varepsilon_{m}\rangle\langle\varepsilon_{n}\vert]\tilde{\rho}_{S}(t)$
are defined by
\begin{equation}
	D[o]\tilde{\rho}_{S}(t)=2o\tilde{\rho}_{S}(t)o^{\dagger}
	-\tilde{\rho}_{S}(t)o^{\dagger}o-o^{\dagger}o\tilde{\rho}_{S}(t)
\end{equation}
for $o=\vert\varepsilon_{n}\rangle\langle\varepsilon_{m}\vert$ or $o^{\dagger}=\vert\varepsilon_{m}\rangle\langle\varepsilon_{n}\vert$. In Eq. (\ref{5}), the decay rates related to the dissipation
channels of the TLS and the bosonic mode are, respectively, defined by
\begin{align}
	\gamma_{\sigma}(\omega_{m,n})=2\pi\varrho_{\sigma}(\omega_{m,n})\lambda^{2}(\omega_{m,n}),\nonumber \\
	\gamma_{a}(\omega_{m,n})=2\pi\varrho_{a}(\omega_{m,n})\eta^{2}(\omega_{m,n}),
\end{align}
where $\varrho_{\sigma}(\omega_{q})$ and $\varrho_{a}(\omega_{k})$
are, respectively, the spectral density functions of the heat baths
associated with the TLS and the bosonic mode, and $\hbar\omega_{m,n}=\varepsilon_{m}-\varepsilon_{n}$
denotes the energy separation between the two eigenstates $\vert\varepsilon_{m}\rangle$
and $\vert\varepsilon_{n}\rangle$ of the QRM.
In our calculations, we suppose that the decay rates $\gamma_{\sigma}(\omega_{m,n})=\gamma_{\sigma}$
and $\gamma_{a}(\omega_{m,n})=\gamma_{a}$, which means that the decay
rates related to all the transitions caused by the same subsystem are identical.
The transition coefficients in Eq.~(\ref{5}) are defined by $\chi_{\sigma,m,n}=\langle\varepsilon_{m}\vert\sigma_{x}\vert\varepsilon_{n}\rangle$ and
$\chi_{a,m,n}=\langle\varepsilon_{m}\vert(a+a^{\dagger})\vert\varepsilon_{n}\rangle$.
In addition, the average thermal-excitation numbers in Eq.~(\ref{5}) are defined
by $\overline{n}_{l=\sigma,a}(\omega_{m,n})=1/[\exp(\hbar\omega_{m,n}/k_{B}T_{l})-1]$,
where $\hbar\omega_{m,n}$ denotes the energy separation between the involved two eigenstates $\vert\varepsilon_{m}\rangle$ and $\vert\varepsilon_{n}\rangle$. The parameters $T_{\sigma}$ and $T_{a}$ are, respectively, the temperatures of the heat baths connected to the TLS and the bosonic mode.

In terms of the transformation $\rho_{S}(t)=e^{-iH_{\mathrm{QRM}}t/\hbar}\tilde{\rho}_{S}(t)\\e^{iH_{\mathrm{QRM}}t/\hbar}$ and quantum master Eq.~(\ref{5}), we can obtain the quantum master equation in the Schr\"{o}dinger picture as
\begin{equation}
	\frac{d}{dt}\rho_{S}(t)=-\frac{i}{\hbar}[H_{\mathrm{QRM}},\rho_{S}(t)]+L_{\mathrm{IHB}}[\rho_{S}(t)],
	\label{8}
\end{equation}
where $\rho_{S}(t)$ is the reduced density matrix of the QRM in the
Schr\"{o}dinger picture, and the dissipator $L_{\mathrm{IHB}}[\rho_{S}(t)]$ is obtained by substituting $\tilde{\rho}_{S}(t)$ with $\rho_{S}(t)$ in Eq.~(\ref{5}). We should point out that the global quantum master equation~(\ref{8}) does not work at the degenerate points in the energy spectrum. This is because we have used the secular approximation in the derivation of the Lindblad dissipators by discarding the related crossing terms. Theoretically, these crossing terms should be kept at the degenerate points. For keeping the uniformity of the systematic description of the Rabi system, we derive the global quantum master equation in the full parameter space, and add this notice to avoid these degenerate points. Note that this notice also works for the Rabi model in the CHB case.

\subsection{Quantum thermalization}
In the non-equilibrium open-system case, the QRM is connected with two heat baths, which could be at different
temperatures. To evaluate the thermalization, we check whether the steady-state density matrix of the QRM can be expressed as a thermal state $\rho_{\mathrm{th}}(T)=Z_{\mathrm{QRM}}^{-1}$\\$\times\exp[-\beta H_{\mathrm{QRM}}]$, where $Z_{\mathrm{QRM}}=\mathrm{Tr}_{\mathrm{QRM}}\{\exp[-\beta H_{\mathrm{QRM}}]\}$
is the partition function of the QRM, with $T$ being the temperature of the thermalized system. In the quantum thermalization process, the environment erases all the initial-state information of the thermalized system.
For the QRM, it is valid when the coupling enters the ultrastrong even deep-strong coupling regimes. Therefore, we should treat the QRM as an effective multiple-level system, i.e., working in the eigenstate representation of the QRM.
As a result, the thermalization of the QRM in the IHB case can be understood as the thermalization of a multiple-level system connected with two heat baths, which would be at either different temperatures or the same temperature. By solving the global quantum master equation (\ref{8}) in the eigenstate representation of the QRM, we find that the steady state of the QRM is a completely mixed state in this representation. Motivated by this feature, we introduce effective temperatures associated with any two eigenstates based on their populations.
If all the effective temperatures are the same, then the steady-state density matrix of the QRM can be written as a thermal state. In this case, we regard it as the quantum thermalization of the QRM.
\begin{figure}
	\begin{centering}
		\includegraphics[width=0.5\textwidth]{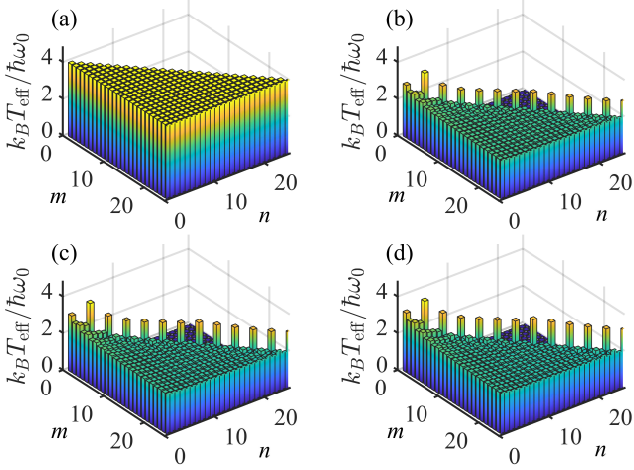}
		\par\end{centering}
	\centering{}\caption{Effective temperatures $k_{B}T_{\mathrm{eff}}/\hbar\omega_{0}$ as functions of the
		energy-level indexes $m$ and $n$ in various cases:
		(a) $k_{B}T_{\mathrm{\sigma}}/\hbar\omega_{0}=k_{B}T_{a}/\hbar\omega_{0}=4$ and $\gamma_{\sigma}/\omega_{0}=\gamma_{a}/\omega_{0}=0.001$;
		(b-d) $k_{B}T_{\mathrm{\sigma}}/\hbar\omega_{0}=2$, $k_{B}T_{a}/\hbar\omega_{0}=4$, $\gamma_{\sigma}/\omega_{0}=0.001$,
		and $\gamma_{a}/\gamma_{\sigma}=1$, $1.5$, $2$ respectively. Other
		parameters used are $\omega_{c}=\omega_{0}$ and $g/\omega_{0}=0.5$.}
	\label{Fig2}
\end{figure}

We denote the populations of the states $\vert\varepsilon_{m}\rangle$
($\vert\varepsilon_{n}\rangle$) as $p_{m}$ ($p_{n}$). Then, we
define the effective temperature
as
\begin{equation}
	T_{\mathrm{eff}}(\omega_{m,n})=(-\hbar\omega_{m,n}/k_{B})\ln\left(\frac{p_{m}}{p_{n}}\right).	
\end{equation}
According to these effective temperatures, we can characterize the thermalization
of the QRM. Based on Eq.~(\ref{8}), we can obtain the equations of motion for the density matrix elements $\langle\varepsilon_{n}\vert\rho_{\mathrm{ss}}\vert\varepsilon_{m}\rangle$. For investigating the thermalization, we focus on the steady state of the system, which can be solved by setting
$\frac{\mathrm{d}}{\mathrm{d}t}\rho_{S}(t)\rightarrow0$, then the steady-state density matrix
elements $\langle\varepsilon_{n}\vert\rho_{\mathrm{ss}}\vert\varepsilon_{m}\rangle$ can be obtained. In terms of the population $p_{n}=\langle\varepsilon_{n}\vert\rho_{\mathrm{ss}}\vert\varepsilon_{n}\rangle$
of the eigenstate $\vert\varepsilon_{n}\rangle$, then the effective
temperatures $T_{\mathrm{eff}}(\omega_{m,n})$ can be calculated accordingly.
In our numerical simulations, we need to truncate the dimension up to $n_{d}$
of the Hilbert space of the bosonic field so that $\sum_{n=1}^{n_{d}}\langle\varepsilon_{n}\vert\rho_{\mathrm{ss}}\vert\varepsilon_{n}\rangle\approx1$.

In Fig.~\ref{Fig2}, we plot the effective temperatures $k_{B}T_{\mathrm{eff}}(\omega_{m,n})\\/\hbar\omega_{0}$ as functions of the energy-level indexes $m$ and $n$ when the scaled bath temperatures  $k_{B}T_{\mathrm{\sigma}}/\hbar\omega_{0}$ and $k_{B}T_{a}/\hbar\omega_{0}$ take various values. Figures \ref{Fig2}(a) and \ref{Fig2}(b)-\ref{Fig2}(d) correspond to the cases of $k_{B}T_{\mathrm{\sigma}}/\hbar\omega_{0}=k_{B}T_{a}/\hbar\omega_{0}$ and $k_{B}T_{\mathrm{\sigma}}/\hbar\omega_{0}\neq k_{B}T_{a}/\hbar\omega_{0}$, respectively. Figure~\ref{Fig2}(a) shows that the QRM can be thermalized under the circumstance of $k_{B}T_{\mathrm{\sigma}}/\hbar\omega_{0}=k_{B}T_{a}/\hbar\omega_{0}$. In this case, the temperature of the thermalized QRM equals to that of the two baths, and the density matrix of the QRM can be written as the thermal state. In particular, the thermalization of the QRM is independent of the nonzero values of the decay rates. In Figs.~\ref{Fig2}(b)-\ref{Fig2}(d), the temperatures of these adjacent energy levels of the QRM are not the same, then the steady state of the QRM cannot be expressed as a thermal state, and hence the QRM cannot be thermalized in this case.

\begin{figure}
	\centering{}\includegraphics[width=0.5\textwidth]{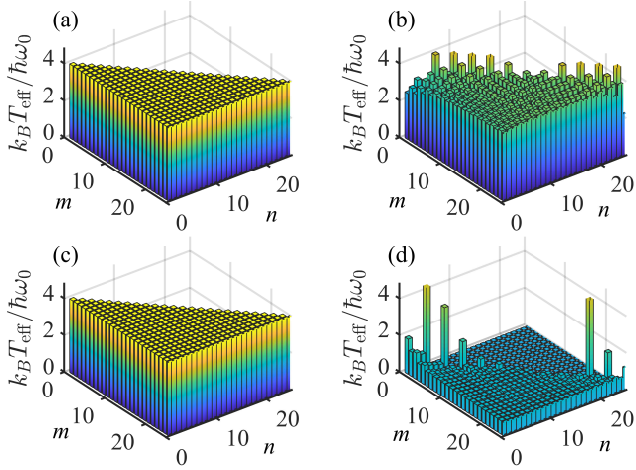}\caption{Effective temperatures $k_{B}T_{\mathrm{eff}}/\hbar\omega_{0}$ as functions of the
		energy-level indexes $m$ and $n$ in various cases: (a) $k_{B}T_{\mathrm{\sigma}}/\hbar\omega_{0}=4$,
		$\gamma_{\sigma}/\omega_{0}=0.001$, and $\gamma_{a}/\omega_{0}=0$;
		(b) $k_{B}T_{\mathrm{\sigma}}/\hbar\omega_{0}=4$, $k_{B}T_{a}/\hbar\omega_{0}=0$, and $\gamma_{\sigma}/\omega_{0}=\gamma_{a}/\omega_{0}=0.001$;
		(c) $k_{B}T_{a}/\hbar\omega_{0}=4$, $\gamma_{a}/\omega_{0}=0.001$, and $\gamma_{\sigma}/\omega_{0}=0$;
		(d) $k_{B}T_{\mathrm{\sigma}}/\hbar\omega_{0}=0$, $k_{B}T_{a}/\hbar\omega_{0}=4$, and $\gamma_{\sigma}/\omega_{0}=\gamma_{a}/\omega_{0}=0.001$.
		Other parameters used are $\omega_{c}=\omega_{0}$ and $g/\omega_{0}=0.5$. }
	\label{Fig3}
\end{figure}

It should be emphasized that the system-bath coupling configuration
is of great importance for the thermalization of the QRM. In the case of
unequal temperatures of the two baths, the QRM cannot be thermalized. When one of the system-bath
couplings is turned off, the thermalization of the QRM will have different results. In Fig.~\ref{Fig3}, we show
the effective temperatures $k_{B}T_{\mathrm{eff}}(\omega_{m,n})/\hbar\omega_{0}$ as functions of the
energy-level indexes $m$ and $n$. Figures~\ref{Fig3}(a) and~\ref{Fig3}(b) correspond to the cases where the bath
of the bosonic mode is decoupled [Fig.~\ref{Fig3}(a)] and the bath
of the bosonic mode is at zero temperature [Fig.~\ref{Fig3}(b)], respectively.
Similar results can also be found when either the TLS decouples from its bath or the bath of the TLS does not
excite the system [Figs.~\ref{Fig3}(c) and~\ref{Fig3}(d)]. These results indicate that the QRM can be thermalized when one of the two baths is decoupled from the QRM.
When the couplings of the QRM with the two baths exist simultaneously, the QRM cannot be thermalized even when one of
the baths is at zero temperature (the temperature of the other bath is not zero).
Consequently, it can be concluded that
the QRM can be thermalized when only one heat bath coupled to the QRM
and cannot be thermalized when the two heat baths are at two different temperatures.

In the above discussions on the thermalization of QRM, we only show the result for the resonant atom-field coupling case. However, we would like to point out that these results concerning the thermalization condition are general. These results also hold for the off-resonant QRM. We have confirmed the thermalization condition by performing the numerical simulations in the off-resonant QRM case. Here, we do not show the off-resonant coupling results for conciseness.

\section{Quantum Thermalization of the QRM in the CHB Case}\label{IV}

In this section, we investigate whether the QRM can be thermalized in the common heat-bath case. Using the similar method as in the IHB case and Eq. (\ref{3}), We can derive a global quantum master equation to govern the evolution of the QRM as
\begin{equation}
	\frac{d}{dt}\rho_{S}(t)=-\frac{i}{\hbar}[H_{\mathrm{QRM}},\rho_{S}(t)]+L_{\mathrm{CHB}}[\rho_{S}(t)],
	\label{10}
\end{equation}
with the dissipator $L_{\mathrm{CHB}}[\rho_{S}(t)]$ given by

\begin{align}
	L_{\mathrm{CHB}}[\rho_{S}(t)]= & \sum_{m,n=1,m>n}^{\infty}\sum_{l=\sigma,a,X}\frac{1}{2}\gamma_{l}(\omega_{m,n})\vert\chi_{l,m,n}\vert^{2}\nonumber \\
	\times & [\overline{n}_{l}(\omega_{m,n})+1]D[\vert\varepsilon_{n}\rangle\langle\varepsilon_{m}\vert]\rho_{S}(t)\nonumber\\
	+ & \sum_{m,n=1,m>n}^{\infty}\sum_{l=\sigma,a,X}\frac{1}{2}\gamma_{l}(\omega_{m,n})\vert\chi_{l,m,n}\vert^{2}\nonumber \\
	\times & \overline{n}_{l}(\omega_{m,n})D[\vert\varepsilon_{m}\rangle\langle\varepsilon_{n}\vert]\rho_{S}(t),
	\label{11}
\end{align}
where the Lindblad superoperator $D[o]$ is defined in Sec.~\ref{III}. We define the
decay rates in Eq. (\ref{11}) as
\begin{align}
	\gamma_{\sigma}(\omega_{m,n})= & 2\pi\varrho_{c}(\omega_{m,n})\lambda^{2}(\omega_{m,n}),  \nonumber\\
	\gamma_{a}(\omega_{m,n})= & 2\pi\varrho_{c}(\omega_{m,n})\eta^{2}(\omega_{m,n}), \nonumber\\
	\gamma_{X}(\omega_{m,n})= & \sqrt{\gamma_{\sigma}(\omega_{m,n})\gamma_{a}(\omega_{m,n})},
\end{align}
which correspond to the dissipations through the TLS, the bosonic
mode, and the cross effect between the two subsystems, respectively.

Similar to the IHB case, we obtain the steady state of quantum master equation (\ref{10}), and find that the density matrix of the QRM in the eigenstate representation is diagonal. Based on the steady-state populations of these eigenstates $\vert\varepsilon_{n}\rangle$, we calculate the effective temperatures assosiated with any two eigenstates $\vert\varepsilon_{m}\rangle$ and $\vert\varepsilon_{n}\rangle$.  We also checked the thermalization of the QRM in the CHB case when the two decay rates take different values. It has been found that the thermalization in the CHB case is independent of the values of the two decay rates (the two decay rates cannot be zero at the same time). Here, we do not present the figure because the results are the same as Fig.~\ref{Fig2}(a). 
We also mention that the thermalizaiton results in the CHB case are also hold for the nonresonant QRM.

\section{Thermal Entanglement in the QRM}\label{V}

In this section, we study thermal entanglement in the QRM. When the QRM is thermalized at temperature
$T$, its density matrix can be written as $\rho_{\mathrm{th}}(T)=\mathrm{\exp}(-\beta H_{\mathrm{QRM}})$\\$/Z_{\mathrm{QRM}}$ with $\beta=1/(k_{B}T)$. We study quantum entanglement
between the TLS and the single-mode bosonic field by calculating the logarithmic negativity of the thermal state.
Note that here we use the logarithmic negativity to quantify the quantum entanglement between the TLS and the bosonic mode because the thermal state is a mixed state. We should point out that, in the zero-temperature case, i.e., $T=0$, the thermal state is reduced to the ground state of the QRM. In some previous papers,\textsuperscript{\cite{bib56,bib57,bib58}} quantum entanglement in the ground state of the QRM has been studied using the von-Neumann entropy. It has been found that the entropy of entanglement monotonically increases and exhibits a saturation effect with the increase of the coupling strength $g/\omega_{0}$ at the resonant case $\omega_{c}=\omega_{0}$. We remind that the logarithmic negativity does not reduce to the entropy of entanglement, and that the logarithmic negativity could be zero even the state is entangled. However, for quantifying quantum entanglement of the QRM in thermal state, the logarithmic negativity is a computable quantity.

For the QRM in the thermal state $\rho_{\mathrm{th}}$$(T)$, its logarithmic negativity
can be calculated by
\begin{equation}
	N=\log_{2}\left|\left|\rho_{\mathrm{th}}^{\mathrm{T_{a}}}(T)\right|\right|_{1},
\end{equation}
where "$\mathrm{T_{a}}$" denotes partial transpose with respect to the bosonic mode. The trace norm is defined by $||\rho_{\mathrm{th}}^{\mathrm{T_{a}}}(T)||_{1}=\mathrm{Tr}\left[\sqrt{(\rho_{\mathrm{th}}^{\mathrm{T_{a}}}(T))^{\dagger}\rho_{\mathrm{th}}^{\mathrm{T_{a}}}(T)}\right]=\sum_{i}\sqrt{\lambda_{i}}$, where $\lambda_{i}$ are the eigenvalues of $\left(\rho_{\mathrm{th}}^{\mathrm{T_{a}}}(T)\right)^{\dagger}\rho_{\mathrm{th}}^{\mathrm{T_{a}}}(T)$. According to the relation $H_{\mathrm{QRM}}\vert\varepsilon_{k}\rangle=\varepsilon_{k}\vert\varepsilon_{k}\rangle$, the thermal state of the QRM can be written as $\rho_{\mathrm{th}}(T)=Z_{\mathrm{{QRM}}}^{-1}\sum_{k=1}^{\infty}\exp(-\beta \varepsilon_{k})\vert\varepsilon_{k}\rangle\text{\ensuremath{\langle}}\varepsilon_{k}\vert$.
For calculation of the logarithmic negativity, we need to express the density matrix $\rho_{\mathrm{th}}$$(T)$ with the bare states $\vert e,m\rangle$ and $\vert g,n\rangle$. Therefore, we expand the eigenstates of the
QRM with the bare basis vectors $\vert e,m\rangle$ and $\vert g,n\rangle$.

Before discussing the thermal entanglement, we first analyze the parameter space of the thermal state, such that the thermal entanglement can be discussed clearly in the full parameter space. For the QRM in the thermal state, its density matrix can be expressed as
\begin{equation}
	\rho_{\mathrm{th}}(T)=Z_{\mathrm{QRM}}^{-1}e^{-\beta\hbar\omega_{0}[\frac{\sigma_{z}}{2}
		+\frac{\omega_{c}}{\omega_{0}}a^{\dagger}a+\frac{g}{\omega_{0}}\sigma_{x}(a^{\dagger}+a)]}.
	\label{13}
\end{equation}

It can be seen from Eq.~(\ref{13}) that the density matrix $\rho_{\mathrm{th}}(T)$ is determined by three ratios: (i)
The scaled inversed temperature $\hbar\omega_{0}\beta$, which can be used to characterize the temperature of the thermalized system. When both the coupling strength $g$ and the resonance frequency $\omega_{c}$ are either smaller than or of the same order of $\omega_{0}$, then the $\omega_{0}$ can be used to characterize the energy scale of the system, and the ratio $\hbar\omega_{0}\beta$ can be used to describe the temperature scale of the system. The relations $\hbar\omega_{0}\beta\gg1$ and $\hbar\omega_{0}\beta\ll1$ stand for the low- and high-temperature limits, respectively. (ii) The frequency ratio
$\omega_{c}/\omega_{0}$, which describes the frequency mismatch between the TLS and the bosonic mode. Here $\omega_{c}/\omega_{0}=1$ and $\omega_{c}/\omega_{0}\neq1$ correspond to the resonant- and nonresonant-couplings between the TLS and the bosonic mode, respectively. (iii) The scaled coupling strength $g/\omega_{0}$, which is the ratio of the coupling strength over the resonance frequency of the TLS. Based on the appoint in ultrastrong coupling, $g/\omega_{0}>0.1$ and $g/\omega_{0}>1$ are considered as the conditions for characterizing the ultrastrong-coupling and deep-strong-coupling regimes of the QRM, respectively.\textsuperscript{\cite{bib17,bib18}}

In the resonant-coupling case $\omega_{c}/\omega_{0}=1$, the density matrix of the thermal state just depends on the two ratios $\hbar\omega_{0}\beta$ and $g/\omega_{0}$. Therefore, we can clearly investigate the dependence of the logarithmic negativity $N$ on the ratios $\hbar\omega_{0}\beta$ and $g/\omega_{0}$. In this way, the dependence of the thermal entanglement in the QRM on the system parameters can be analyzed clearly. In the nonresonant-coupling case, we can investigate the dependence of the thermal entanglement on the two ratios $\omega_{c}/\omega_{0}$ and $g/\omega_{0}$ when the ratio $\hbar\omega_{0}\beta$  takes different values.

\begin{figure}[bthp]
	\begin{centering}
		\includegraphics[width=0.47\textwidth]{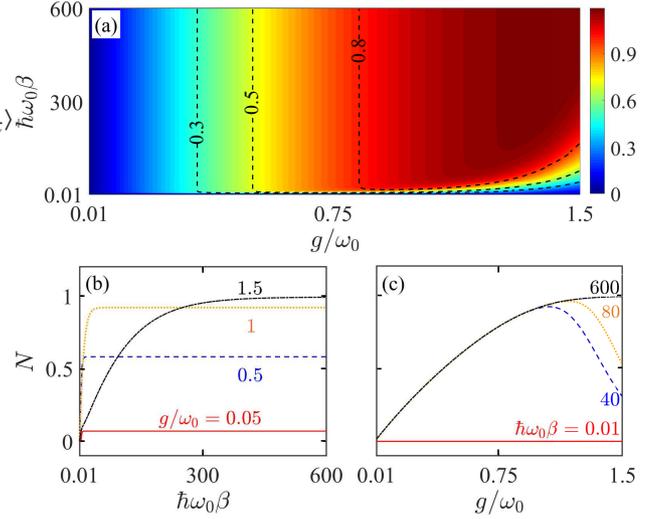}
		\caption{(a) Logarithmic negativity $N$ of the QRM in the
			thermal state versus $\hbar\omega_{0}\beta$ and $g/\omega_{0}$. (b) Logarithmic negativity $N$ as a function of $\hbar\omega_{0}\beta$, when $g/\omega_{0}=0.05$, $0.5$, $1$,
			and $1.5$. (c) Logarithmic negativity $N$ as a function of $g/\omega_{0}$ at $\hbar\omega_{0}\beta=0.01$, $40$, $80$, and $600$. Other parameter used is $\omega_{c}=\omega_{0}$.}
		\label{Fig6}
		\par\end{centering}
\end{figure}
In Fig.~\ref{Fig6}, we plot the logarithmic negativity $N$ of the QRM
in the thermal state as a function of $\hbar\omega_{0}\beta$ and $g/\omega_{0}$ in the resonant case $\omega_{0}=\omega_{c}$.
Here the thermal state $\rho_{\mathrm{th}}(T)$ just depends on the two ratios $\hbar\omega_{0}\beta$ and $g/\omega_{0}$. Therefore,
Fig.~\ref{Fig6}(a) shows the contour map of the thermal entanglement in the overall view. In a macroscopic view, thermal entanglement
shows a wedge-ridge pattern. The peak value of the entanglement is approximately located around the coupling $g/\omega_{0}\approx1$.
In addition, we can see that the thermal entanglement disappears in the high-temperature limit $\hbar\omega_{0}\beta\ll1$.
In the low-temperature limit $\hbar\omega_{0}\beta\gg1$, the thermal entanglement approaches a stable value, which is the ground state entanglement of the QRM. This point can be seen more clearly in panel (b).
Here, we see that for a given $g/\omega_{0}$, the thermal entanglement increases with the increase of $\hbar\omega_{0}\beta$, and then the thermal entanglement tends to be a stable value in strong, ultrastrong, and deep-strong coupling regimes. We note that this feature can be understood by analyzing the dependence of the eigenstate population $p_{\vert\varepsilon_{m}\rangle}$ on the $\hbar\omega_{0}\beta$. In addition, we can see that the stable value of the logarithmic negativity $N$ is in a monotonous order with respect to the coupling strength $g/\omega_{0}$. In particular, for a larger coupling strength $g/\omega_{0}$, the value of $\hbar\omega_{0}\beta$ corresponding to the turning point is larger.

\begin{figure}[htbp]
	\begin{centering}
		\includegraphics[width=0.48\textwidth]{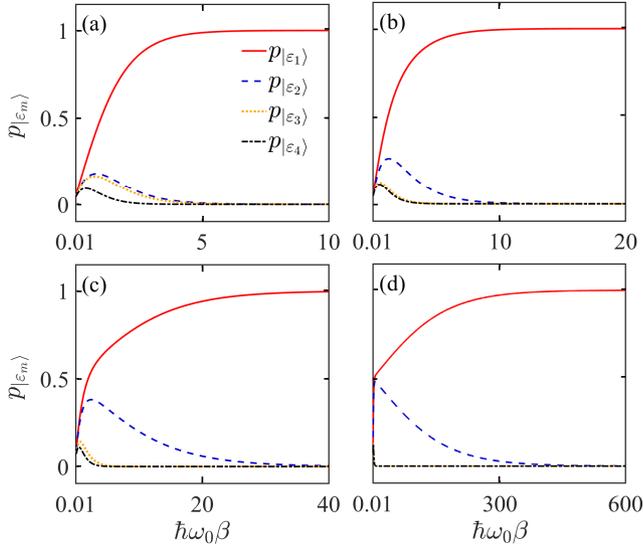}
		\caption{Eigenstate populations $p_{\vert\varepsilon_{m}\rangle}$ for $m=1-4$ as functions of $\hbar\omega_{0}\beta$ at different values of $g/\omega_{0}$: (a) $g/\omega_{0}=0.05$, (b) $g/\omega_{0}=0.5$, (c) $g/\omega_{0}=1$, and (d) $g/\omega_{0}=1.5$. Here $\omega_{c}=\omega_{0}$.}
		\label{Fig8}
		\par\end{centering}	
\end{figure}
To see the dependence of the logarithmic negativity $N$ on the coupling strength, in Fig.~\ref{Fig6}(c) we plot the $N$ as a function of $g/\omega_{0}$ at different values of $\hbar\omega_{0}\beta$.
Here we can see that, for $\hbar\omega_{0}\beta=0.01$, the system is in the high-temperature limit, and hence the thermal engtanglement disappears. For $\hbar\omega_{0}\beta=600$, the $N$ increases with the increase of $g/\omega_{0}$. When $\hbar\omega_{0}\beta=40$ and $80$, however, we find that the $N$ experiences a nonmonochromatic change, it increases firstly and then decreases. These features can be explained by analyzing the dependence of the populations of these eigenstates on the value of $\hbar\omega_{0}\beta$.

In Fig.~\ref{Fig8}, we show the populations $p_{\vert\varepsilon_{m}\rangle}$ as functions of $\hbar\omega_{0}\beta$ at different values of $g/\omega_{0}$. Here, we can see that, for a given value of $g/\omega_{0}$, the ground state probability $p_{\vert\varepsilon_{1}\rangle}$ increases with the increase of the inverse temperature $\hbar\omega_{0}\beta$. Accordingly, the populations $p_{\vert\varepsilon_{m}\rangle}$ for $m=2$, $3$, $4$ decrease with the increase of $\hbar\omega_{0}\beta$. For a larger value of $g/\omega_{0}$, a larger value of $\hbar\omega_{0}\beta$ is needed to make sure the system reaching its ground state. For example, $\hbar\omega_{0}\beta \approx 600$ is needed to make sure the system reaching the ground state for $g/\omega_{0}=1.5$ as shown in Fig.~\ref{Fig8}(d). For the cases of $\hbar\omega_{0}\beta=40$ and $80$ in Fig.~\ref{Fig6}(c), when the coupling strength $g/\omega_{0}$ is larger than a value around $1$, the system will be in a mixed state involving the ground state and the first excited state. Here, the ground state $\vert\varepsilon_{1}\rangle$ and the first excited state $\vert\varepsilon_{2}\rangle$ are near degenerate, as shown in Fig.~\ref{Fig14}(b), and hence the entanglement is not a monochromatic function of the coupling strength. For the case of $\hbar\omega_{0}\beta=600$, when the coupling strength changes from $g/\omega_{0}=0.01$ to $1.5$, the system is always in the ground state, then the engtanglement in the ground state increases with the increase of the coupling strength, this feature matches the result when the entanglement is measured by the von Neumann entropy.\textsuperscript{\cite{bib56,bib57,bib58}}

\begin{figure}
	\begin{centering}
		\includegraphics[width=0.48\textwidth]{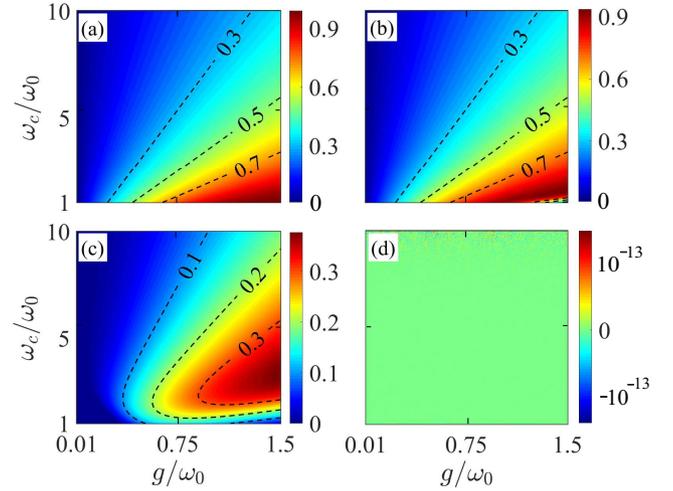}
		\caption{Logarithmic negativity $N$ of the QRM in the
			thermal state versus the parameters $\omega_{c}/\omega_{0}$ and $g/\omega_{0}$. Other parameters used in (a)-(d) are $\hbar\omega_{0}\beta=600$, $30$, $2$, and $0.01$, respectively.}
		\label{Fig7}
		\par\end{centering}	
\end{figure}
In Fig.~\ref{Fig7}, we plot the logarithmic negativity $N$ of the QRM
in the thermal state as a function of the ratios $\omega_{c}/\omega_{0}$ and $g/\omega_{0}$, when the bath temperature takes different values. From Figs.~\ref{Fig7}(a) to \ref{Fig7}(d), we take $\hbar\omega_{0}\beta=600$, $30$, $2$, and $0.01$, corresponding to a gradual change of the thermalized temperature from the low-temperature limit to the high-temperature limit. From Fig.~\ref{Fig7}, we can see three features of the thermal entanglement. (i) Thermal entanglement exists in the low-temperature and finit-temperature cases, and there is no thermal entanglement in the high-temperature limit. This result is understandable because the thermal noise is harmful to quantum effect, and hence the quantum entanglement disappears in the high-temperature limit. In the zero-temperature limit, the thermal equilibrium state is reduced to the ground state, and the ground-state entanglement in the QRM has been studied in some previous works.\textsuperscript{\cite{bib56,bib57,bib58,bib59}} In the high-temperature limit, the thermal state becomes a completly mixed state, i.e., the density operator in the eigenstate representation becomes an identity matrix. Therefore, the thermal entanglement disappears in the high-temperature limit. (ii) In the finit-temperature case, the thermal entanglement exhibits a wedge-ridge pattern. For a given value of $\omega_{c}/\omega_{0}$, the logarithmic negativity $N$ increases with increase of the coupling strength $g/\omega_{0}$. For a given coupling $g/\omega_{0}$, the logarithmic negativity $N$ first increases and then decreases when the ratio $\omega_{c}/\omega_{0}$ changes from $1$ to a much larger value. (iii) At a relatively low temperature, the peak value of the logarithmic negativity $N$ appears approximately around the resonance-coupling case, i.e., $\omega_{c}/\omega_{0}\approx1$. In particular, the slope of the ridge line corresponding to the peak value of entanglement increases slightly with the increase of the coupling strength $g/\omega_{0}$. For a higher thermalized temperature, the ridge line moves slightly towards the increasing direction of the value $\omega_{c}/\omega_{0}$.

\section{Discussions on the experimental implementation}\label{VI}
In this section, we present some discussions on the experimental implementation of this scheme. Concretely, we analyze the experimental feasibility based on four aspects. (i) The realization of the QRM with practical systems. The physical mode considered in this paper is general, and it can be realized in realistic physical setups with which the QRM can be implemented. Currently, the QRM can be realized in many systems, including various cavity-QED systems,\textsuperscript{\cite{bib16,bib60}} circuit-QED systems,\textsuperscript{\cite{bib61}} and other hybrid systems\textsuperscript{\cite{bib62,bib63}} described by the interaction between a two-level system (artificial atoms, spins, charge states etc.) and a single-mode bosonic field (cavity field, mechanical resonator, LC resonator etc.). (ii) The parameter condition of the physical model. In this work, the ultrastrong- and deep-strong-coupling regimes are considered. Therefore, the candidate physical system should enter these two regimes. Currently, both the ultrastrong and deep-strong couplings have been realized in many physical platforms.\textsuperscript{\cite{bib17,bib18}} (iii) The control of the effective bath temperatures. To study the thermalization, the candidate systems should be chosen such that the effective bath temperature can be tuned to control the equilibrium and nonequilibrium environments. Experimentally, the control of the effective bath temperature can be realized in circuit-QED systems.\textsuperscript{\cite{bib64}} (iv) The measurement of the eigenstate populations and eigenstate entanglement. In realistical physical systems, the measurement of the two-level system and the bosonic mode can be realized. However, it remains unexplored for measuring the population and entanglement of the eigenstates for the QRM. Therefore, to experimentally implement the present scheme, the experimental measurement of the eigenstate population and entanglement need to be developed. Based on the above analyses, we can conclude that the present scheme should be experimentally accessible once the experimental measurement of the eigenstate population and entanglement can be realized. Thanks to the recent great advances in ultrastrong couplings, much effort is devoting to the study of the measurement of the ultrastrongly coupled systems in the eigenstate representation. This will provide a hopeful prospect for the experimental realization of this scheme.

\section{Conclusion}\label{VII}
In conclusion, we have studied the quantum thermalization of the open QRM, which is connected
with either two IHBs or a CHB. We have derived the global quantum master
equations in the eigenstate representation of the QRM to govern the evolution of the QRM in both the IHB and CHB cases. Based on the steady-state populations of the eigenstates, we have studied the quantum thermalization of the QRM by checking whether all the effective temperatures between any two eigenstates are the same. In
the IHB case, we have found that, when the two IHBs have the
same temperature (different temperatures), the QRM can (cannot) be
thermalized. When one of the two IHBs is decoupled from the QRM, then
the QRM can be thermalized with the coupled bath. In the CHB case,
the system can always be thermalized. Additionally, we have studied
quantum entanglement between the TLS and the bosonic field when
the QRM is in a thermal state. The logarithmic negativity for the thermal state of the QRM has been calculated
in both the resonant- and nonresonant-coupling cases. The dependence of the thermal entanglement on the system parameters and the thermalized temperature has been found and discussed in detail. This work will give an insight to quantum information processing at finite temperature.

\section*{Acknowledgments}
J.-F.H. is supported in part by the National Natural Science Foundation of China (Grant No. 12075083),
and Natural Science Foundation of Hunan Province, China (Grant No. 2020JJ5345). J.-Q.L. was supported in part by National Natural Science Foundation of China (Grants No. 12175061 and No. 11935006) and the Science and Technology Innovation Program of Hunan Province (Grants No. 2021RC4029, No. 2017XK2018, and No. 2020RC4047).

\section*{Conflict of interest}
The authors declare no conflicts of interest.


\end{document}